\documentclass[twocolumn,showpacs,amsmath,amssymb,floatfix]{revtex4}
\usepackage{graphicx}
\usepackage{dcolumn}
\usepackage{bm}
\newcommand{\beq}{\begin{equation}}
\newcommand{\eeq}{\end{equation}}
\newcommand{\bea}{\begin{eqnarray}}
\newcommand{\eea}{\end{eqnarray}}

\begin{document}
\author{D. Voitenkov}
\affiliation{Institute for Physics and Power Engineering, 249033
Obninsk, Russia}
\author{S. Kamerdzhiev}
\affiliation{Institute for Physics and Power Engineering, 249033
Obninsk, Russia}
\author{S. Krewald}\affiliation{Institut fuer Kernphysik, Forschungszentrum Juelich,
D-52425 Juelich, Germany}
\author{E. E. Saperstein}
\affiliation{Kurchatov Institute, 123182 Moscow}
\author{S. V. Tolokonnikov}
\affiliation{Kurchatov Institute, 123182 Moscow} \affiliation{Moscow
Institute of Physics and Technology, 123098 Moscow, Russia}
\title{Self-consistent calculations of quadrupole moments of the first 2$^+$ states \\
in Sn and Pb isotopes}


\begin{abstract}
A  method of calculating  static moments of  excited states and
transitions between excited states is formulated for non-magic
nuclei within the Green function formalism. For these
characteristics, it leads to a noticeable difference from  the
standard QRPA approach. Quadrupole moments of the first 2$^+$ states
in Sn and Pb isotopes are calculated using the self-consistent TFFS
based on the Energy Density Functional by Fayans et al. with the set
of parameters DF3-a fixed previously. A reasonable agreement with
available experimental data is obtained.

\end{abstract}

\maketitle

\section{Introduction}
To reliably predict properties of unstable nuclei at the modern
level of microscopic nuclear theory at least two conditions  need to
be fulfilled. First, it is necessary to take into account the
single-particle continuum which is especially important for the
description of  nuclei with small separation energies.
 Second, an approach should be used with the self-consistency relation between
 the mean field and effective interaction. This makes it possible to
  use only one set of parameters 
 instead of two sets, for the effective interaction and  mean field,
 in non-self-consistent approaches.
  The self-consistency improved noticeably
 the predictive power of the theory even  on the RPA or QRPA level,
  for the review see \cite{reviewPaar}.

 Nowadays, it is also necessary to add to these conditions the accounting for
 phonon coupling (PC). This problem has been  studied  for a long time within several
 approaches which are referred as the Quasiparticle-Phonon model (QPM) \cite{solov},
 (Q)RPA+PC \cite{colo1994}, Extended Theory of Finite Fermi Systems
  (ETFFS) \cite{revKST} (non-selfconsistent
  cases) and the self-consistent versions,  (Q)RPA+PC \cite{sarchi} and
  the ETFFS in the Quasiparticle Time Blocking Approximation \cite{tsel2007}
(ETFFS(QTBA)) \cite{avd2007}.  For magic and semi-magic nuclei, such
approaches are based on the fact that in these  nuclei there is a
small parameter $g^2$, the dimensionless square of the phonon
creation amplitude. For brevity, we call this weak PC approximation
where only the $g^2$ terms are taken into account the  $g^2$
approximation.

   \begin{figure}[ht!]
\includegraphics[scale=0.4]{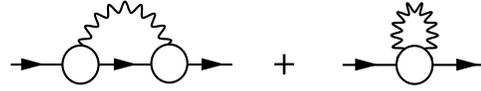}
\vspace{4mm} \caption{$g^2$ order corrections to the mass operator
in magic nuclei. The circles with one wavy line in the first term
are the phonon creation amplitudes. The second term is the phonon
tadpole.}
\end{figure}

 In the framework of the Green function (GF) formalism,
all the conditions under discussion have been realized and have
shown their importance for stable nuclei too 
\cite{revKST,avd2007,tsel2007}. However, all the above mentioned
approaches  dealing with the PC did not take into account
\textit{all} the $g^2$ terms, thus  limiting themselves with the pole diagrams
only, see the first diagram in Fig. 1 where diagrams for the mass
operator are displayed. The second diagram represents the sum of all
the non-pole diagrams that we call the phonon tadpole now.

 The problem of consistent consideration of all the $g^2$
 terms including tadpoles was analyzed in the  article by Khodel  \cite{khodel1976}
 on the base of the general self-consistency relations for finite Fermi systems \cite{FKh}.
The method developed was practically realized for magic nuclei,
mainly for static nuclear characteristics, within  the
self-consistent TFFS \cite{sapkhodel1982}. It was found
 that, as a rule, the tadpole contributions in magic nuclei
 are noticeable and are often of opposite sign as compared with those of the
 pole terms. The first attempts to include the tadpole effects for nuclei with pairing
 and for consideration of dynamical problems  were  recently made in \cite{kaevsap} and
 \cite{kaevavevoit2011}, respectively.

In the PC problem, the $g^2$ corrections to the mean field shown in Fig.1 
 have been mainly studied  up to now \cite{sapkhodel1982}. They are, as a rule,
 smaller than the corresponding mean field values  and could be partially hidden in the phenomenological
parameters used.
 In this work, within the GF method, we concentrate our attention on   more
 delicate characteristics  which are proportional to  $g^2$ themselves.
Namely, we analyze the static moments of excited states
 and transitions  between excited states. To ensure the self-consistency, we
 use the self-consistent TFFS based on the Energy Density Functional by Fayans et al.
 \cite{fayans2000} with the DF3-a set of parameters fixed previously
 \cite{Tol-Sap}.
  We briefly consider
some static and low-energy characteristics in magic  and generalize the method for
non-magic even-even
 nuclei (Sects. II and III, respectively).
 Within this approach, and using some  approximations
 we  perform the first self-consistent  calculations of
  static moments of the first $2^+$ excited states in
 even-even tin and lead isotopes (Sect.IV)

 The quadrupole moments of excited states
  have been calculated earlier within QPM in \cite{vdovin1,vdovin2}.
 In Ref.\cite{broglia1972}, the authors performed the calculations, which are
 similar to ours in Sect.IV, using the method later called  Nuclear Field Theory
 and a phenomenological approach with the set of phenomenological parameters taken
 from experiment  for each  nucleus. In \cite{broglia1972}, a reasonable agreement 
 was obtained with the
 experimental data for Sn  and Ni isotopes when available at that time.
  The main difference of our approach from \cite{broglia1972,vdovin1,vdovin2}
  is its full self-consistency on the (Q)RPA level and absence of any
  phenomenological or fitted parameters. These features
  will allow characteristics of unstable nuclei to be calculated.

\section{Magic nuclei}

To describe the PC effects in magic nuclei with the consistent
account of all the $g^2$ terms, we follow  the method by Khodel
\cite{khodel1976}. In the $g^2$ approximation, the matrix element
$M_{LL}$ for a static moment of the excited state (phonon) with the
orbital angular moments $L$ in a static external field $V^{0}$, is
determined in terms of the change of  the one-particle GF in the
field of this phonon:
\begin{equation}
\label{matrixelem} M_{LL} = \int V^{0}(\textbf{r})\delta^{(2)}_{LL}
G(\textbf{r},\textbf{r},\varepsilon )d\textbf{r}
\frac{d\varepsilon}{2 \pi  \imath} ,
\end{equation}
where \bea \label{deltaLL} \delta^{(2)}_{L L} G =\delta_{L}(G g_{L}
G) = G(\varepsilon) g_{L}
G(\varepsilon +  \omega_{L}) g_{L} G(\varepsilon) \\
 \nonumber +G(
\varepsilon) g_{L} G( \varepsilon - \omega_{L}) g_{L} G(
\varepsilon)+ G( \varepsilon) \delta_L g_{L} G( \varepsilon), \eea
where $g_L$ is the amplitude for the production of the L phonon
with the  energy $\omega_L$ and
$ \delta_L g_{L}$, the variation of $g_L$ in the  field of other $L$
phonon. Substituting Eq. (2) into Eq. (1), we obtain in the symbolic
form:
\begin{equation}
\label{matrixelem4}
M_{L L} = V^0 G g_L G g_{L} G + V^0 G G \delta_L g_{L},
\end{equation}
which is illustrated in Fig.2 where the blocks $g_L$
 and $ \delta_L g_{L}$ enter.

It is convenient to transform this expression in such a way that the
effective field $V$ will appear instead of the external field
$V^{0}$. They are connected with the TFFS equation, \cite{AB} \beq V
= V^0+{\cal F} A  V, \label{Vef_s} \eeq where $ {\cal F}$ is the
effective particle-hole (ph) interaction and $A$ is the ph
propagator (the integral over energy of the product of two
single-particle GF's). After regrouping the terms in Eq.(3) and
iterating the integral equation for the quantity $ \delta_L g_L $ in
the effective interaction $ {\cal F}$  (for  details, see
\cite{sapkhodel1982,kaevavevoit2011}), we obtain the ultimate
expression,
\begin{equation}
\label{matrixelem2} M_{L L}= VGg_LGg_LG +VGG \delta_L {\cal F} GG
g_{L},
\end{equation}
which is illustrated in Fig. 3.
 It contains now the effective field $V$, instead of  $V^{0}$
 in Eq. 3, and the quantity $\delta_L {\cal F}$ in the second term which
 denotes  the variation of the effective ph interaction  ${\cal F}$ in
 the field of the $L$ phonon. For the density dependent TFFS
 effective  interaction ${\cal F}(\rho)$, the following ansatz can
 be readily obtained \cite{khodel1976,sapkhodel1982}:
\begin{equation}
\label{deltaLF} \delta_L {\cal F}({\bf r})= \frac{\partial {\cal
F}}{\partial \rho}  \rho_L^{\rm tr}(r) Y_{LM}(\bf n),
\end{equation} where $\rho_L^{\rm tr}=Ag_L$ is the transition
density for the $L$ phonon excitation. The first term of Eq.(5)
coincides with the result by Speth \cite{speth1970} while the second
one with the $ \delta_L {\cal F} $ quantity is a generalization  to
take into account all the $g^2$ terms.

\begin{figure}[h]
\includegraphics[scale=0.4]{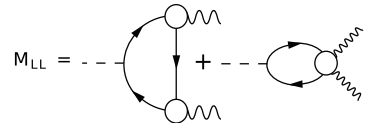}
\caption{Matrix element $M_{L L}$ for magic nuclei, Eq.
(\ref{matrixelem4}).}
\end{figure}

\begin{figure}[h]
\includegraphics[scale=0.4]{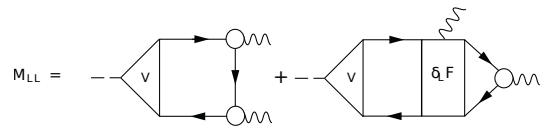}
\caption{Matrix element  $M_{L L}$ in the form of Eq.
(\ref{matrixelem2}).}
\end{figure}

All the above equations can be readily modified for such processes
as the transition between the excited states $L$ and $L'$ in the
external field $V^0(\omega = \omega_{L'}- \omega_L)$ or the
excitation of the two-phonon state $L+L'$ in the external field
$V^0(\omega = \omega_{L'}+\omega_L)$. The static moments case corresponds 
to $\omega = 0$.

\section{Non-magic nuclei. Comparison with QRPA.}
In the case of nuclei with pairing, it is necessary to use four GF's
($G,G^h,F^{(1)},F^{(2)}$ in the usual notation \cite{AB}). To
describe phonons,  one has to use the complete set of the QRPA
equations which include the ph, hp, pp, and hh
 channels and four effective fields $V$, $V^h$, $d^{(1)}$ and $d^{(2)}$ \cite{AB}, respectively.
 As the pp and hh channels  give a small contribution in the case of the first 2$^+$ levels \cite{tolokn2011},
 which is considered in the next Section,  we
 do not consider these channels and, accordingly, the fields $d^{(1)}$ and $d^{(2)}$. Then we obtain
 eight terms $M^{(i)}_{L L^\prime}$ instead of one in Eq.(5). The typical two terms , $M^{(1)}$
  and $M^{(5)}$ are shown in Fig.4.
 In this study we will  consider the case without the terms with $\delta_{L} {\cal F}$ and
 $\delta_{L} {\cal F}^{\xi}$,
  see Sect.IV

The integral of the three GF's  $A^{(1) {\rm pair}}_{123}$ has the form
\begin{eqnarray}
\label{A1} &A^{(1) {\rm pair}}_{123}(\omega_L, \omega_{L^ \prime})
 = \int \! G_1(\varepsilon) G_2(\varepsilon +
\omega_L) G_3(\varepsilon + \omega_{L^\prime}) \displaystyle\frac{d
\varepsilon}{2 \pi \imath}\nonumber \\\noalign{\smallskip} &\!\!=
\displaystyle\frac{u^2_1 u^2_2 v^2_3}{(\omega_L +
E_{13})(\omega_{L^\prime} + E_{23})} + \displaystyle\frac{v^2_1
v^2_2 u^2_3}{(\omega_L - E_{13}) (\omega_{L^\prime} -
E_{23})}\nonumber \\\noalign{\smallskip} &+
\displaystyle\frac{1}{\omega + E_{12}} \left( \frac{u^2_1 v^2_2
u^2_3}{E_{23} - \omega_{L^\prime}} - \frac{u^2_1 v^2_2 u^2_3}{E_{13}
+ \omega_L}\right) \\\noalign{\smallskip} & +
\displaystyle\frac{1}{\omega - E_{12}} \left( \frac{v^2_1 u^2_2
v^2_3}{E_{23} + \omega_{L^\prime}} - \frac{v^2_1 u^2_2 u^2_3}{E_{13}
- \omega_L} \right),\nonumber
\end{eqnarray}
where $E_{12} = E_1 + E_2,  E_1 = \sqrt{(\varepsilon_1 - \mu )^2 +
\Delta_1^2}$ and the low index 1 = $(n_1, l_1, j_1)$ (spherical
nuclei) is the set of single-particle quantum numbers.

\begin{figure*}
\includegraphics[scale=0.9]{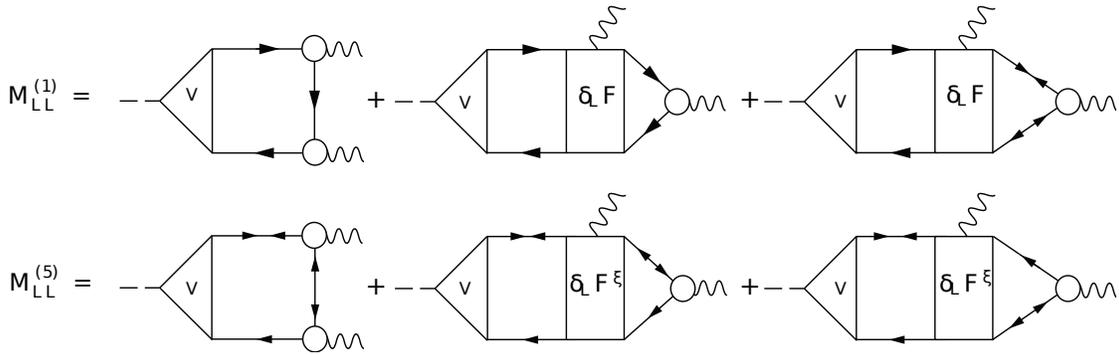}
\caption{Matrix elements for  $M^{(1)}_{L L}$ and $M^{(5)}_{L L}$
for non-magic nuclei.}
\end{figure*}

In the pairing case there are eight such integrals of three GF's
$A^{(i){\rm pair}}_{123}$, where $i$ = 1-8. After a long algebra,
one can obtain the final formula for the diagonal matrix element
$M_{LL}$ with L = L$^\prime$, which corresponds to the case of
static  quadrupole moment of the excited L state. Just  this case is
considered in the next section .

\begin{eqnarray}
M_{LL} = \sum_{123} (-1)^{M_{L}+1}
\begin{pmatrix}
I &L &L\\
0 &L &-L
\end{pmatrix}
\left\lbrace\begin{matrix}
I &L &L\\
j_3 &j_2 &j_1
\end{matrix}\right\rbrace \nonumber\\
 \times\!<\!1\!\parallel V \parallel \!2\!> <\!3\!\parallel g_L \parallel\! 1\!>
  <\!2\!\parallel g_L \parallel \!3\!> \!\sum_{i=1}^8 A^{(i){\rm pair}}_{123}\,,
\end{eqnarray}
where
\begin{eqnarray}
\label{sumA} &\sum\limits_{i=1}^8 A^{(i){\rm pair}}_{123} \nonumber
\\ \noalign{\smallskip} & = \displaystyle\left( \frac{1}{(\omega_L +
E_{13})(\omega_L + E_{23})} + \frac{1}{(\omega_L - E_{13})(\omega_L
- E_{23})} \right)\nonumber\\\noalign{\smallskip}
&\times\displaystyle\left[ u^2_1 u^2_2 v^2_3 - v^2_1 v^2_2 u^2_3
 + \frac{\Delta_1 \Delta_2 } {4 E_1 E_2} (u^2_3 - v^2_3)\right. \nonumber\\ \noalign{\smallskip}
&+\displaystyle \left.\frac{\Delta_1 \Delta_3}{4 E_1 E_3} (u^2_2 -
v^2_2) + \frac{\Delta_2 \Delta_3}{4 E_2 E_3} (u^2_1 - v^2_1)\right]+
\\\noalign{\smallskip} &\displaystyle\frac{1}{E_{12}}\!\!
\displaystyle\left[\frac{2E_{23}( u^2_1 u^2_3 v^2_2\! - v^2_1 v^2_3
u^2_2)}{E_{23}^{2}-\omega_L^{2}}\right.
{+}\displaystyle\frac{2E_{13}( u^2_2 u^2_3 v^2_1\! - v^2_3 v^2_2
u^2_1)}{E_{13}^{2}-\omega_L^{2}} \nonumber\\ \noalign{\smallskip} &-
\displaystyle\left(\frac{ \Delta_1 \Delta_2 } {2 E_1 E_2} (u^2_3 -
v^2_3) + \frac{\Delta_1 \Delta_3}{2 E_1 E_3} (u^2_2 - v^2_2)\right.
\nonumber\\ \noalign{\smallskip} &+\left.\displaystyle\frac{\Delta_2
\Delta_3}{2E_2 E_3} (u^2_1 - v^2_1) \right)
\displaystyle\left.\left(\frac{E_{13}}{E_{13}^{2}-\omega_L^{2}}
 +\frac{E_{23}}{E_{23}^{2}-\omega_L^{2}}\right)\right].\nonumber
\end{eqnarray}

Let us  compare this  expression with the respective result of  QRPA approach.
Here we mean  the usual way which uses the QRPA wave functions for
the matrix element between two excited states.
In Ref.\cite{ponomarev1998} the expression for the matrix element
(the B(E2) quantity, to be exact) has been derived using  the
bare external  field and, like in our case,  the QRPA wave
functions without the pp and hh-channels. 
The first square brackets (in the first half of Eq.(9)) coincide completely 
with the factor $v^{-}_{12}u^{+}_{23}u^{+}_{31}$ in
Refs.\cite{ponomarev1998,solov}.
Thus, the first half of Eq.(9) corresponds to
the expression $v^{-}_{12}(\psi_{23}\psi_{31} + \phi_{23}\phi_{31})$
in \cite{ponomarev1998}
because the phonon amplitudes $\psi$ and $\phi$ contain, by
definition, the denominators $(E_{12}-\omega_L)$ and $(E_{12} +
\omega_L)$, respectively. Therefore, the second half of Eq.(9)
(with the  common factor 1/$E_{12}$), generalizes the usual QRPA approach.
This part of Eq.(9)
describes the contribution of the ground state correlations (the
so-called graphs going back) to our ``triangle'' with the integral
of three GF's, see Eq. (7). We
will calculate the quantitative contribution of such correlations in
the next section. The second generalization is the appearance of the
effective field V, which depends on the frequency $\omega = \omega_L \pm
\omega_L^\prime $  instead of external field $V^0$, which does not
depend on the frequency.  The terms with $\delta_L {\cal F}$ and
$\delta_L {\cal F}^\xi $ are the third generalization of the QRPA
approach

\section{Calculations of static quadrupole moments of the first $2^+$ states in tin and
lead isotopes}

 The quadrupole moment of the excited state L is connected with the matrix element $M_{LL}$,
 Eq. (8) ($I = 2,\; V({\bf r}) = V(r)Y_{20}({\bf n})$ ), as follows:
\begin{equation}
 Q = \sqrt{\frac{16\pi}{5}} M_{LL}.
 \end{equation}

In the recent work \cite{KVarXiv},
 we calculated the static quadrupole moment of the first 3$^-$ level in $^{208}$Pb,
 using only the first term in Eq. (5) (or Fig. 3),
 and obtained the value $Q_{\rm theor}$ = -- 0.26 b which is in a reasonable agreement with
 the experimental one,
 $Q_{\rm exp}$ = -- 0.34$\pm$0.15 b \cite{stone}. At present, we have calculated the second term with
$\delta_{L} {\cal F}$ from Eq. (6) and found that it gives
approximately 1\% as compared with the first term.
Though the contribution of the terms with $\delta_{L} {\cal F}$ and
$\delta_{L} {\cal F}^\xi $
 in non-magic nuclei should be analyzed specially, it  can hardly be
considerable in the problem under consideration and is omitted in
the present calculations.

  We calculated the quadrupole moments of the first 2$^+$ states in
  non-magic tin and lead isotopes according to Eqs. (8 -- 10)
  in the $\lambda$-representation with
  self-consistent single-particle wave functions $\phi_{\lambda}$ obtained within the EDF method
  \cite{fayans2000} with the functional DF3-a
  \cite{Tol-Sap}.
  The spherical box
  of the radius $R{=}16\;$fm was used to simulate the single-particle continuum. We examined the 
  dependence of calculation results on the cut-off energy $E_{\rm max}$
  and have found that the value of $E_{\rm max}{=}100\;$MeV ensures 1\% accuracy.  To calculate
  the quantities V and $g_L$, the  results of our previous article \cite{tolokn2011} have been used where
  all the calculations were performed in the coordinate representation using the same self-consistent
  DF3-a basis as in the present calculation of the matrix element  $M_{LL}$.
  Thus, no fitted
parameters was used in the present calculations.

  The results are given in Table 1 and Figs. 1, 2. Except for the case of $^{112}$Sn 
  and $^{208}$Pb nuclei,
  we obtained a reasonable agreement with experimental data
  \cite{stone}.
   Unfortunately, they have rather big errors so that we need some better measurements to check
   our approach.

   We have also calculated the contribution of the ground state correlations term in
Eq. (9) (see the discussion at the end of Sect. III) and obtained
that it is rather considerable.  For some nuclei  it is about
60\% of all the triangle contribution. This very interesting result
will  be discussed  in more detail separately.

\begin{table}[t]
\caption{Quadrupole moments $Q\;$(e\;b) of the first 2$^+$ states in
Sn and Pb isotopes.
 ($Q^{n}$ and $Q^{p}$ are the neutron and proton contributions to
the final result  $Q_{\rm tot}{=} Q^{n}$ + $Q^{p}$.
)}

\begin{tabular}{l c c c c}
\hline \hline \noalign{\smallskip}
 nucl.  & $Q^{\rm n}$
&\hspace*{1.ex} $Q^{\rm
p}$\hspace*{1.ex} &\hspace*{1.ex}$Q_{\rm tot}$\hspace*{1.ex} &\hspace*{1.5ex} $Q_{\rm exp}$ \cite{stone}\\
\noalign{\smallskip}
\hline
\noalign{\smallskip}
$^{102}$Sn &-0.05   &-0.01   &-0.07 &--\\
$^{104}$Sn &-0.18   &-0.03   &-0.21 &--\\
$^{106}$Sn &-0.28   &-0.05   &-0.33 &--\\
$^{108}$Sn &-0.31   &-0.07   &-0.38 &--\\
$^{110}$Sn &-0.38   &-0.10   &-0.48 &--\\
$^{112}$Sn &-0.32   &-0.11   &-0.43 &-0.03(11)\\
$^{114}$Sn &-0.15   &-0.11   &-0.26 &0.32(3), 0.36(4)\\
$^{116}$Sn & 0.00   &-0.10   &-0.10 &-0.17(4), +0.08(8)\\
$^{118}$Sn & 0.10   &-0.09   & 0.01 &-0.05(14)\\
$^{120}$Sn & 0.12   &-0.08   & 0.05 &+0.022(10), -0.05(10)\\
$^{122}$Sn & 0.09   &-0.07   & 0.02 &-0.28 $<Q<$+0.14\\
$^{124}$Sn & 0.00   &-0.06   &-0.06 &0.0(2)\\
$^{126}$Sn &-0.08   &-0.05   &-0.12 &--\\
$^{128}$Sn &-0.10   &-0.03   &-0.13 &--\\
$^{130}$Sn &-0.05   &-0.01   &-0.07 &--\\
$^{132}$Sn & 0.04   & 0.00   & 0.04 &--\\
$^{134}$Sn & 0.00   &-0.01   &-0.01 &--\\
$^{190}$Pb &-0.60   &-0.29   &-0.89 &--\\
$^{192}$Pb &-0.77   &-0.35   &-1.12 &--\\
$^{194}$Pb &-0.90   &-0.39   &-1.28 &--\\
$^{196}$Pb &-0.85   &-0.38   &-1.23 &--\\
$^{198}$Pb &-0.67   &-0.35   &-1.02 &--\\
$^{200}$Pb &-0.27   &-0.23   &-0.50 &--\\
$^{202}$Pb & 0.02   &-0.15   &-0.13 &--\\
$^{204}$Pb & 0.18   &-0.07   & 0.11 &+0.23(9)\\
$^{206}$Pb & 0.11   &-0.02   & 0.10 &+0.05(9)\\
$^{208}$Pb & 0.01   & 0.04   & 0.05 &-0.7(3)\\
\\
\hline \hline
\end{tabular}
\end{table}

\begin{figure}[ht!]
\includegraphics[scale=0.3]{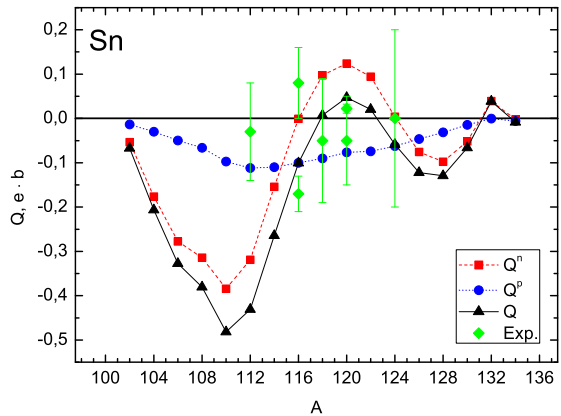}
\caption{(Color online) Quadrupole moments of the first 2$+$ excited
states in even Sn isotopes.}
\end{figure}

\begin{figure}[t!]
\includegraphics[scale=0.3]{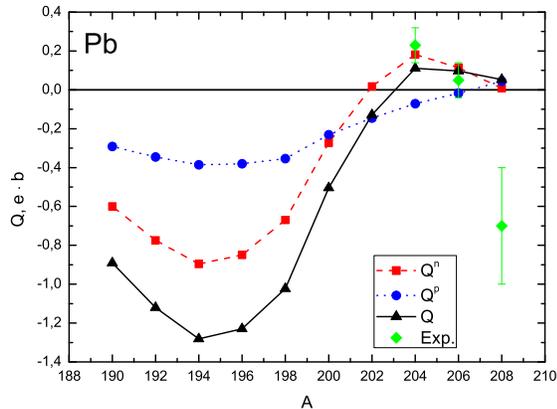}
\caption{(Color online) Same as in Fig.5 but for Pb isotopes}
\end{figure}

\section{Conclusion}
We have considered the method to calculate  static moments of
the excited states and transitions between excited states, which,
generally speaking, are
described within the QRPA, with taking all the g$^2$ terms into account
in magic and non magic nuclei. It was shown that, in addition to the
old results \cite{speth1970,broglia1972}, new terms  with
$\delta_{L}F$ and $\delta_{L}F^{\xi}$ appear, which  contain the density
dependence of both the ph and pp effective interactions.
 We have performed the self-consistent calculations of the
static quadrupole moments of the first 2$^+$ states for Sn and Pb
isotopes using the known functional's parameters set DF3-a.
Except for the $^{112}$Sn and $^{208}$Pb cases, a reasonable  agreement has been obtained 
with the experiment available, though, as a rule, the experimental data have big 
errors.  
 In these calculations
we did not take into account  new terms with $\delta_{L}F$ and
$\delta_{L}F^{\xi}$ .
As all modern
microscopic calculations deal with the density dependence
of ph and pp effective interactions, it is important to analyze  their
 role in the problems under consideration. This rather
complicated problem will be considered separately.

\section{Acknowledgment}
 Four of us, S. T., S. Ka., E. S., and D. V.,
are grateful to Institut fuer Kernphysik, Forschungszentrum Juelich
for hospitality. The work was partly supported by the DFG and RFBR
Grants Nos.436RUS113/994/0-1 and 09-02-91352NNIO-a, by the Grants
NSh-7235.2010.2  and 2.1.1/4540 of the Russian Ministry for Science
and Education, and by the RFBR grants 09-02-01284-a, 11-02-00467-a.


\newpage

\begin{references}
\bibitem{reviewPaar}
Nils Paar, Dario Vretenar, Elias Khan and Gianluca Colo, Rep. Prog.
Phys. \textbf{70}, 691 (2007).

\bibitem{solov} V. G. Soloviev, {\it Theory of Complex Niclei},
(Oxford: Pergamon Press, 1976).
\bibitem{colo1994}
G. Colo, Nguyen Van Giai, P. F. Bortignon, R. A. Broglia, Phys. Rev.
C \textbf{50},1496 (1994).
\bibitem{revKST}
S. Kamerdzhiev, J. Speth, G. Tertychny, Phys. Rep. \textbf{393}, 1 (2004).
\bibitem{sarchi}
D. Sarchi, P. F. Bortignon, G. Colo, Phys. Lett. \textbf{ B601}, 27
(2004).
\bibitem{tsel2007}
V. Tselyaev, Phys. Rev. C \textbf{75}, 024306 (2007).
\bibitem{avd2007}
A. Avdeenkov, F. Gruemmer, S. Kamerdzhiev {\it et al.}, Phys. Lett.
\textbf{B653},196 (2007).

\bibitem{khodel1976}
V. A. Khodel, Sov. J. Nucl. Phys. \textbf{24}, 376 (1976).
\bibitem{FKh}
S. A. Fayans, V. A. Khodel, JETP Lett. \textbf{17}, 633, (1973).
\bibitem{sapkhodel1982}
 V. A. Khodel and E. E. Saperstein, Phys. Rep. {\bf 92}, 183 (1982).

\bibitem{kaevsap}
S. Kamerdzhiev, E. E. Saperstein, Eur. Phys. J. \textbf{ A37}, 159
(2008).
\bibitem{kaevavevoit2011}
S. P. Kamerdzhiev, A. V. Avdeenkov, D. A. Voitenkov, Phys. Atom.
Nucl. \textbf{74}, 1478 (2011).

\bibitem{fayans2000}
S. A. Fayans, S. V. Tolokonnikov, E. L. Trykov, and D. Zawischa,
Nucl. Phys.  {\bf A676}, 49 (2000).
\bibitem{Tol-Sap} S. V. Tolokonnikov and E. E. Saperstein, Phys. Atom. Nucl. {\bf 73}, 1684 (2010).

\bibitem{vdovin1}
 A. I. Vdovin, Ch. Stoyanov,  Izv. Akad. Nauk SSSR, Ser. Fiz., \textbf{38}, 2598 (1974).
\bibitem{vdovin2}
 A. I. Vdovin, Ch. Stoyanov,  Izv. Akad. Nauk SSSR, Ser.  Fiz.,\textbf{ 38}, 2604 (1974).
\bibitem{broglia1972}
R. A. Broglia, R. Liotta and V. Paar, Phys. Lett. \textbf{38B}, 480
(1972).

\bibitem{AB} A. B. Migdal, {\it Theory of finite Fermi systems and applications to
atomic nuclei} (Wiley, New York, 1967).

\bibitem{speth1970}
J. Speth, Z. Phys. \textbf{ 239}, 249 (1970).
\bibitem{tolokn2011}
S. V. Tolokonnikov, S. Kamerdzhiev, D. Voitenkov, S. Krewald, E. E.
Saperstein, arXiv:1107.4232v2[nucl-th],  Phys.
Rev. C 84, 064324 (2011).
\bibitem{ponomarev1998}
V. Yu. Ponomarev, Ch. Stoyanov, N. Tsoneva, M. Grinberg, Nucl. Phys.
\textbf{A635}, 470 (1998).
\bibitem{KVarXiv}
S. Kamerdzhiev, D. Voitenkov, arXiv: 1110.0654 [nucl-th] (2011).
\bibitem{stone}
N. J. Stone, Atomic Data Nuclear Table \textbf{90}, 75 (2005).
\end{references}
\end{document}